\documentclass[sigconf]{acmart}

\AtBeginDocument{%
  }

\setcopyright{acmlicensed}
\copyrightyear{2018}
\acmYear{2018}
\acmDOI{XXXXXXX.XXXXXXX}
\acmConference[Conference acronym 'XX]{Make sure to enter the correct
  conference title from your rights confirmation email}{June 03--05,
  2018}{Woodstock, NY}
\acmISBN{978-1-4503-XXXX-X/2018/06}

\usepackage{tikz}
\usepackage{caption}
\usepackage{blindtext}
\usepackage{tcolorbox}
\usepackage[final]{pdfpages}
\usepackage{lipsum,multicol}
\usepackage{xcolor}
\usepackage{tikz}
\usepackage{listings}
\usepackage{enumitem}
\usepackage{hyperref}
\usepackage{amsfonts}
\usepackage{wrapfig}
\usepackage{subcaption} 
\usepackage{adjustbox}
\usepackage{colortbl}
\usepackage{fancybox}
\usepackage{multirow}
\usepackage[normalem]{ulem}
\useunder{\uline}{\ul}{}
\usepackage{enumitem}

\usepackage{booktabs}

\newcommand{\ie}{\textit{i.e.,}\xspace}
\newcommand{\eg}{\textit{e.g.,}\xspace}

\newtcolorbox{boxK}{
    fontupper = \small,
    sharpish corners, 
    boxrule = 0pt,
    toprule = 0pt, 
}

\newcommand*\circled[1]{\tikz[baseline=(char.base)]{
            \node[shape=circle,draw,inner sep=0.5pt] (char) {#1};}}

\newcommand{\sm}{\textit{SM}\xspace}
\newcommand{\alife}{A\textit{Life}\xspace}
\newcommand{\sdlc}{\textit{SDLC}\xspace}

\newcommand{\llms}{LLMs\xspace}

\newcommand{\approach}{$\Psi$\textit{-Arch}\xspace}

\newtheorem{defn}{Definition}[section]

\begin{document}
\title{ 
Towards Enabling An Artificial Self-Construction Software Life-cycle via Autopoietic Architectures
}

\author{Daniel Rodriguez-Cardenas}
\email{dhrodriguezcar@wm.edu}
\orcid{1234-5678-9012}
\affiliation{%
  \institution{William \& Mary}
  \city{Williamsburg}
  \state{VA}
  \country{USA}
}

\author{David N. Palacio}
\email{davidnad@microsoft.com}
\affiliation{%
  \institution{Microsoft}
  \city{Redmont}
  \state{WA}
  \country{USA}
}

\author{Denys Poshyvanyk}
\email{dposhyvanyk@wm.edu}
\affiliation{%
  \institution{William \& Mary}
  \city{Williamsburg}
  \state{VA}
  \country{USA}
}

\renewcommand{\shortauthors}{Rodriguez-Cardenas et al.}

\begin{abstract}
Software engineering research has focused on automating maintenance and evolution processes to reduce costs and improve reliability. The emergence of \textbf{foundation models (FMs)} with strong code understanding and reasoning abilities offers new opportunities for autonomous software behavior. Inspired by Artificial Life (\textit{ALife}), we propose a fundamental shift in the Software Development Life-Cycle (\textit{SDLC}) by introducing \textit{self-construction} mechanisms that enable software to evolve and maintain autonomously. This position paper explores the potential of \textit{Autopoietic Architectures}, specifically $\Psi$-\textit{Arch}, as a foundational framework for self-constructing software. We first analyze the limitations of traditional maintenance approaches and identify gaps in current SDLC automation. Subsequently, we outline the core challenges in achieving self-construction, including integrating foundation-model-based reasoning units and establishing novel architectural paradigms. Although this paper does not present a definitive solution, it seeks to catalyze discourse and inspire research toward a new paradigm in software engineering—one in which self-constructing software represents the next frontier in SDLC automation.
\end{abstract}





\maketitle

\section{Introduction}
\label{sec:introduction}






Software Development Life-Cycle (\sdlc) outlines five phases to produce a software artifact \ie that is designed, developed, tested, deployed and maintained \cite{10.1145/2543581.2543595,Lehman1980ProgramsEvolution}. Specifically, the software maintenance phase is the most critical and costly in the \sdlc\cite{10.1007/978-981-16-0739-4_28,895984}. Software maintenance entails \textit{any modifications made to a system after deployment}. In fact, these modifications involve the knowledge of the practitioners \cite{10.1145/3597503.3623330} to design, construct, and improve the source code. Recent software maintenance (\sm) research has focused on automating repetitive tasks using Machine Learning~ (ML) and, more recently, foundational models (FMs) - large-scale pre-trained models trained on vast datasets of code and natural language with broad capabilities across diverse tasks.

Automating software maintenance ensures that software systems remain efficient, reliable, adaptable, and relevant over time \cite{9568959,10548798}. Several studies have investigated software maintenance automation on downstream tasks such as testing \cite{gruber2023automatictestgenerationtools,10765036}, and program repair \cite{10172693,10.1145/3597503.3623310,SequenceR,SANER2019}. Additionally, foundation model-based code generation has been extensively studied in the field of software maintenance~\cite{Ciniselli2021AnES, ICSE2021Transformers,Tufano2018AnTranslation,Wang2017Semantics-awareCode,BinghamBrown2017TheMutants,White2016DeepDetection} to assist practitioners with large language models (\llms) demonstrating remarkable abilities to understand code semantics~\cite{merrill_can_2024}, generate patches~\cite{jin_inferfix_2023, ramanan_aspire_2024}, and think about software behavior~\cite{wei2025swerladvancingllmreasoning}. However, as software products grow in size and complexity, \textit{canonical} ML-assisted maintenance becomes increasingly challenging and error-prone \cite{10.1145/3597503.3639095}, powered by the reasoning and generative capabilities of foundation models~\cite{wu_reasoning_2024}.


\begin{figure}[h]
 \centering
    \includegraphics[width=\linewidth]{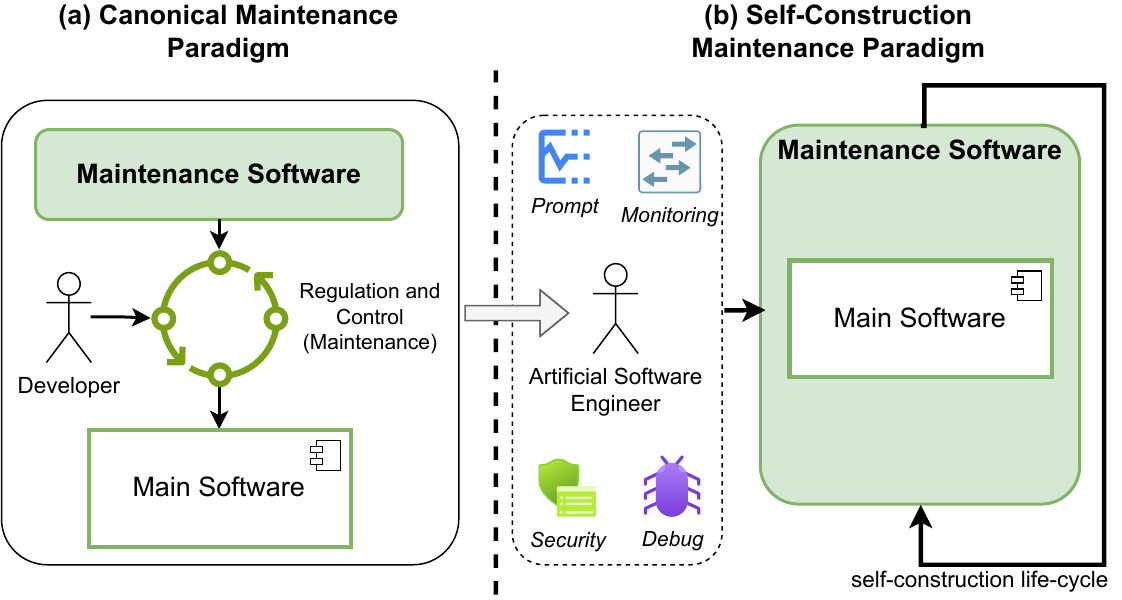}
     \caption{\small{Maintenance Paradigm Shift: a) Software Maintenance (\sm) is independent from the main software, and b) \sm is wrapping the main software.}} 
     \label{fig:newD}
\end{figure}

This canonical vision of software maintenance has permeated research methods in the software community, leaving no room for alternative approaches due to the remarkably successful use of statistical learning technologies such as transformers \cite{vaswani2023attentionneed} or reinforcement learning \cite{wei2025swerladvancingllmreasoning}. Irrespective of the exponentially growing canonical research on maintenance, a significant gap remains as this \sm automation has not yet been addressed from the perspective of autonomous construction. This autonomous mechanism is achieved by incorporating artificial life (\alife) elements to enable \textbf{autopoietic} features (\eg self-organization, self-assembly, or self-replication) for \textit{self-construction}~\cite{Sayama_2024,Dorin2024Alife,arcas2024computationallifewellformedselfreplicating}, powered by the reasoning and generative capabilities of fundation models.

\alife is an interdisciplinary field that examines life-like properties—such as self-organization, adaptation, and evolution—in natural and artificial systems \cite{Dorin2024Alife}. Integrating \alife principles with foundation model capabilities enables maintenance frameworks that not only perform predefined tasks, but also autonomously adapt and evolve under changing conditions. We propose self-construction maintenance as a replacement for the canonical maintenance paradigm within the \sdlc. This paper challenges the traditional understanding of \sm and introduces an autonomous mechanism that supports the software construction life-cycle.


This paper introduces an \textbf{autopoietic architecture} called \approach for the artificial self-construction software life-cycle. \approach is an extended and adapted version of von Neumann's self-replicating model \cite{VonNeumann1966}. We present a first draft for the formal notation of \approach and a complete description of an additional self-construction reasoning unit $\Gamma$. \approach establishes the foundation for next-generation \sdlc through \textit{artificial software engineering}.

In this paper, we explain our \approach and how to integrate an autopoietic architecture in a software system using \llms. We highlight the necessity of using causal reasoning to help the system make decisions on self-refactoring, where FMs provide the generative and reasoning capabilities required for autonomous decision-making.




\section{What are the Autopietic Architectures for?}\label{sec:impact}
We assert that the software maintenance research conducted on automatic techniques to date will be easily integrated into the autopoietic architecture (\approach). Consider the following scenario; a software engineer wants to eliminate a useless functionality in a software system $S$. The useless functionality can be mapped to sections of the source code that are executed, but its results are never required. This is a typical example of \textit{dead functionality} elimination. This engineer wants the system $S$ to self-identify the impact of the elimination of functionality and remove additional dead code. At this point, we can notice that the system $S$ includes the FM-powered solution, in other words, $S$ is an FM by itself. Using an autopoietic architecture (\approach), the system $S$ is controlled by a reasoning unit $\Gamma$ that decides how to react in the presence of dead code and create a refactoring. The autopoietic architecture automatically generates a new, refactored replica of itself to replace the actual system, ensuring that the replica runs stably.

We envision a formal and nature-inspired approach that artificially enables the \textit{self-construction} of the Software Engineering Life-Cycle using \approach. \autoref{fig:newD} illustrates the paradigm shift from a canonical ML-driven to a self-construction maintenance technique. This self-construction paradigm allows software systems to create better replicas of themselves at run-time, leaveraging the code understanding, generation, and reasoning capabilities.
\section{Background \& Context}
\label{sec:background}





Foundation models are large-scale pre-trained models trained on broad datasets that can be adapted to a wide range of tasks~\cite{marl-book}. Unlike traditional Machine Learning models that require task-specific training, FMs leverage transfer learning and self-supervised learning to develop general-purpose capabilities that can be fine-tuned for specific applications. In the context of software maintenance, FMs such as GPT-4, CodeLlama, and StarCoder have demonstrated remarkable capabilities in code generation~\cite{dou_stepcoder_2024}, program repair~\cite{jin_inferfix_2023,10.1145/3597503.3623310}, vulnerability detection~\cite{shimmi_ai-based_2025,rahman_towards_2024}, and code comprehension~\cite{velasco_toward_2025}. 

The term \textit{autopoiesis} refers to a process by which systems continuously regenerate and sustain their components and connections autonomously \cite{MaturanaHumberto1980AutopoesisCognition,Bedau2007ArtificialLife}. Under this autopoietic automation vision, software systems are perceived as dynamic entities with inherent self-regulation, adaptability, and, in particular, \textit{self-construction} behaviors. 

The autopoiesis concept becomes feasible as the recent research on FMs, particularly \llms for SE and agentAI4SE~\cite{yang2024sweagentagentcomputerinterfacesenable,arora2024masaimodulararchitecturesoftwareengineering,zhang2024surveylargelanguagemodels,shen_shortcutsbench_2025} has demonstrated the potential to cover the required components for an autopoietic architecture. This perspective transcends conventional ML techniques by enabling a paradigm where software architectures incorporate self-maintenance, self-repair, and autonomous evolution. As a result, autopoietic principles provide a foundation for \textbf{next-generation artificial software engineering}, fostering resilient, adaptive, and self-sustaining computational ecosystems and redefining the SDLC.

The Software Development Life-Cycle (\sdlc) enables a software product by involving technological resources and stakeholders in a well-designed pipeline. This pipeline consists of several phases, including planning, analysis, design, development, testing, deployment, and \textit{maintenance}. Each phase has its own set of activities and deliverables that feed into the next phase. Software Maintenance (\sm), the last phase, is the longest and, therefore, the most costly because it stays active as long as the software remains in production\cite{10.1007/978-981-16-0739-4_28,895984}. Considerable research attention has been devoted to \sm on the automation of specific \textit{tasks}. These tasks assist maintenance activities such as clone detection \cite{White2016DeepDetection}, bug-fixing \cite{Tufano:ASE18, Tufano:MSR18}, impact analysis, traceability \cite{9284034}, refactoring\cite{Nader-palacio2018AssessingDetection,Mkaouer2016AOpportunities}, bad smell detection \cite{Romano2018ACode,Tufano2017WhenAway}, and vulnerability detection\cite{10.1145/3597503.3623345, 10.1145/3597503.3639173, 10764854}. 

It is generally accepted that \sm automation operates using machine learning (ML) approaches or evolutionary computation (EC) solutions (\ie Search-Based Software Engineering \cite{10.1145/2379776.2379787}). Consequently, several maintenance tasks, within the Software Development Life-Cycle \sdlc, still require the active intervention of practitioners. For example, automating \textit{bug-fixing} with a statistical learning approach depends upon collecting, filtering, deduplicating, and transforming code-based datasets, tasks generally performed by practitioners. Despite the breadth and depth of existing research for automating maintenance tasks, none of those ML and EC approaches have been addressed from the point of view of a \textit{self-adaptive technology}. 

Self-adaptivity comprises the ability of a software system to assess and regulate its behavior\cite{Macias-Escriva2013Self-adaptiveApplications}. Recent progress in self-adaptivity investigates and develops sophisticated systems such as multi-agent software \cite{marl-book}, component-based software engineering \cite{10481576}, configurable systems \cite{ye2025distilledlifelongselfadaptationconfigurable}, and model-driven architectures \cite{10608368}. These systems can adapt their behavior in response to environmental changes (\eg new requirements, limited resources, or code changes). Remarkably, this adaptive capability has also been further studied in the context of Artificial Life (\alife) -particularly- with the introduction, definition, and formalization of \textbf{autopoietic systems}~\cite{Dorin2024Alife}.

\section{From Canonical to Self-Construction Life-Cycle}
\label{sec:position}

\begin{figure*}[ht]
\centering
\caption{Autopoietic Architecture \approach to enable artificial self-construction software life-cycle}
    \includegraphics[width=0.95\textwidth]{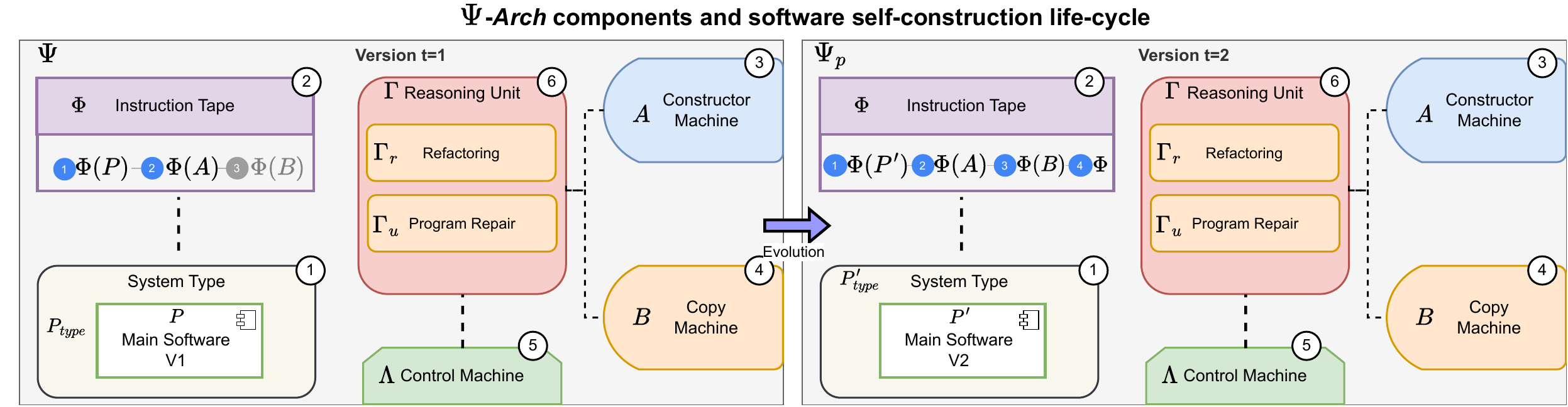}

     \label{fig:arch}
     \vspace{-1.2em}
\end{figure*}
\begin{table*}[!h]
\caption{The Reasoning Unit $\Gamma$ Depends Upon The Maintenance Task}
\vspace{-1.2em}
\label{table:reasoning}
\centering
\scalebox{0.75}{

\def\arraystretch{1}\tabcolsep=10pt
\begin{tabular}{p{0.13\linewidth}p{0.32\linewidth}p{0.32\linewidth}p{0.32\linewidth}}
\multicolumn{1}{c}{\textbf{Dimension}} & \multicolumn{1}{c}{\textit{Observation}} & \multicolumn{1}{c}{\textit{Intervention}} & \multicolumn{1}{c}{\textit{Retrospection}} \\ \hline
\textbf{Representation} & Probability $p(M|\epsilon)$ & Causal Effect $do\{p(M|\epsilon)\}$ & Counterfactuals $E{[}Y_M|M,Y{]}$ \\
\textbf{Purpose} & Take decisions about maintenance tasks based on observations & To intervene in the architecture to assess the causal effect of a new replica & To imagine possible scenarios where a modification  on the architecture takes place \\
\textbf{Source of Inf.} & Quality SE Metrics $\Pi$ & Traceability Links $L$, SE Metrics $\Pi$ & Depends upon the maintenance task $M$ \\
\textbf{Algorithm} & Variational Inference & Front/Back door adjustment & Front/Back door adjustment \\
\textbf{Agent Question} & What if the unit sees a metric $M$ under desired thresholds? & If the unit modifies a sub-component $P$ in the main system to fulfill a maintenance task $M$, will enhance a $\Pi$ metric? & What if the unit had assessed one possible maintenance solution $M$ that was not being considered before? \\ \hline
\end{tabular}%
}
\vspace{-1.2em}
\end{table*}

Following Asimov's vision about a self-replicating machine~\cite{Asimov1956LastQuestion}, this paper examines the relationship between the premise \textit{software systems can self-replicate into a ``better, decentralized, and intelligent'' version of themselves} and the Software Development Life-Cycle (\sdlc). This self-replication process embodies the \textit{self-construction} of adapted internal mechanisms, which compose the original software, to attend an \textit{environment} of functional (\ie what the system should do) and non-functional (\ie how the system should perform) requirements.

Assuming the autopoietic automation vision, we must recognize and acknowledge the feasibility of a new type of autonomous architecture that leverages an artificial self-construction life-cycle and the FM's ability to understand code semantics and generate syntactically correct, functionally appropriate patches~\cite{wei2025swerladvancingllmreasoning}. We refer to this autonomous mechanism as \textit{Autopoietic Architectures} (\approach). To attain an artificial self-construction life-cycle paradigm, beyond the canonical perspective (see \autoref{fig:newD}-a), we argue that \approach can be used as a self-construction standard to implement systems whose \textit{self-replicating components} are calibrated and well communicated. This self-construction paradigm draws inspiration from the Von Neumann kinetic beast \cite{VonNeumann1951,VonNeumann1966} \footnote{Langton later enhanced this foundational model by introducing a simpler self-replicating automaton to address the problem of evolutionary capabilities \cite{Langton1989ALife}}. 

The \approach embodies six self-replicating components: the main (or original) software, an instruction tape, three machines (\eg constructor, copy and control machines) and a reasoning unit. The self-replicating process starts when the control machine $\Lambda$ causes the copy machine $B$ to replicate the instruction tape $\Phi$.


The \textit{first} component encapsulates the original software to replicate $P$ (\autoref{fig:arch}-\circled{1}). $P_{type}$ refers to Lehman's categorization of software systems~\cite{Lehman1980ProgramsEvolution} that encompasses specified systems ($S$-Systems), problem-solving systems ($P$-Systems), and evolving systems ($E$-systems). The \textit{second} component (\autoref{fig:arch}-\circled{2}) consists of an \textit{instruction tape}, $\Phi$, which stores the meta-data of \approach. This meta-data contains detailed instructions on constructing the entire architecture from its source code. The instruction tape $\Phi$ could incorporate advanced, goal-oriented techniques for system maintenance (\ie automated refactoring, patching, repairing, and re-engineering processes). The \textit{third} component (\autoref{fig:arch}-\circled{3}) reflects a \textit{constructor machine}, denoted as $A$, which is responsible for reading any instruction $\phi$ from the instruction tape $\Phi$ to construct components. The \textit{fourth} component is a \textit{copy machine} $B$ (\autoref{fig:arch}-\circled{4}). $B$ can replicate $\Phi$ instruction, \ie $B$ can copy $\Phi$ meta-data -including $\Phi(B)$. The \textit{fifth} component is the control machine $\Lambda$ (\autoref{fig:arch}-\circled{5}). $\Lambda$ produces an isolated copy of $\Psi$ and communicates with the \textit{reasoning unit} $\Gamma$ (\autoref{fig:arch}-\circled{6}). $\Gamma$ is the \textit{sixth} and last component oracle that formulates causal questions to make decisions~\cite{Tucci2013IntroductionDo-Calculus}.

\begin{defn}
The copy process generates a $\Phi$ final instruction tape  
\begin{equation} 
B + \Phi(\Psi) \rightarrow \Phi(\Psi)
\end{equation}
where the operand `$+$' is the use of two components ($B$ and $\Phi$) and the operand `$\rightarrow$' is a realization or copy. In other words, $B$ takes the meta-data instructions and generates a copy of  $\Phi(\Psi)$. Define $\Delta = A+B$. Thus, $\Psi$ would be the $\Delta + P$ composition. Replacing in the previous formula:

\begin{equation} 
B + \Phi(\Delta + P) \rightarrow \Phi(\Delta + P)
\end{equation}
\end{defn}

\begin{defn}
Now, $\Lambda$ causes A to construct the components described by the instruction in $\Phi(\Psi)$.  
The construction process generates a copy of $\Psi$ 
\begin{equation} 
A + \Phi(\Psi) \rightarrow \Psi
\end{equation}
\end{defn}


\begin{defn}
The control machine $\Lambda$ synchronizes the machines correctly to produce the tape $\Phi$ and the $\Psi$-Architecture by adding the copy of $\Phi(\Psi)$ to the new $\Psi$
\begin{equation} 
A + B + \Gamma + \Phi(\Psi) \rightarrow \Psi + \Phi(\Psi)
\end{equation}

Now, let's include in the tape machine the information of the main software or $P$, the reasoning unit $\Gamma$, and rename the entire architecture as $\Psi_P$, therefore $\Delta= A + B + \Gamma$: 
\begin{equation} 
 \Psi_P = \Delta + \Phi(\Delta+P) = A + B + \Gamma + \Phi(A + B + \Gamma + P)
\end{equation}
so we can observe that  $\Psi_P \rightarrow A + B + \Gamma + P + \Phi(A + B + \Gamma + P)$

which means:
\begin{equation} 
\Psi_P \rightarrow \Psi_P + P
\end{equation}
\end{defn}

Finally, after following previous definitions, we obtain the original software $P$, with the respective optimizations $P'$, and its instruction tape. $P$ is embedded in the autopoietic architecture \approach. Note that $P'$ optimizations originate from the reasoning unit $\Gamma$. $\Gamma$ introduces small instructions to the instruction tape so $\Phi(P)$ evolves to $P'$. $\Gamma$ can include other complementary maintainability tasks \eg performing control tasks, such as software traceability.

This reasoning component $\Gamma$ is the most challenging to design as it comprises unique and necessary knowledge to address self-construction tasks. However, recent research has demonstrated the reasoning capabilities of FMs via prompt engineering and agent orchestration~\cite{toran_autogpt_2023}. To properly define this reasoning unit, we decompose it, as detailed in \autoref{table:reasoning}, into five key dimensions (\eg component representation, purpose, source of information, algorithm employed, and agent question). 

In addition to those dimensions, we can classify reasoning tasks into three levels according to Pearl's causality approach: observation, intervention, and retrospection \cite{Pearl2009CausalOverview}. Firstly, the goal of the \textit{observation} level is to predict a maintenance task $M$ based on given evidence $\epsilon$. The prediction is a conditional distribution $p(M|\epsilon)$ where $M$ can be corrective $M_c$, adaptive $M_a$, perfective $M_p$, or preventive $M_v$ activity. Secondly, the \textit{intervention} level aims to measure the causal effect when the reasoning unit $\Gamma$ modifies a sub-component in $P$ to accomplish a maintenance task $M$. Finally, the \textit{retrospection} level aims to estimate the expected value of hypothetical conditions. For example, consider the reasoning unit $\Gamma$ performed an adaptive maintenance task $M_a$ with an observed result $Y=y$, but now the reasoning unit wants to assess what would have happened ($Y=?$) if, instead of $M_a$, the unit had performed $M_v$. The previous situation formally defines a counterfactual prediction. The reasoning unit can measure this hypothetical condition using the expected value $E(Y_{M=M_v} | M=M_a, Y=y)$.

\section{Discussion \& Implications}
\label{sec:discussion}



The emergence of self-evolving and autonomous FM agents represents a natural evolution beyond the autopoietic architecture (\approach) framework presented in this work. Recent research demonstrates that FM-powered agents can transcend static, human-designed configurations to become adaptive systems capable of continuous self-improvement through interaction data and environmental feedback. These developments align closely with the self-construction paradigm proposed in \approach and suggest promising directions for extending the autopoietic software life-cycle.

Because the reasoning unit automates some maintenance tasks, a replica of the autopoietic architecture \approach serves as an enhanced copy of the original (\ie main) software. This enhancement formally represents what software researchers refer to as \textit{software evolution}\cite{10.1145/2593882.2593893}. In light of our autopoietic architecture \approach, the proposed work seeks to explore the following research questions (RQs):

\begin{itemize}[leftmargin=*, labelwidth=0pt, labelindent=0pt]
	\item $RQ_1$ \textit{How effective is using \alife elements combined with foundation models to create software systems that exhibit self-construction behavior?} The effectiveness of \alife to create software systems that exhibit living features (\ie self-replication) has been widely studied \cite{Bedau2007ArtificialLife}. However, we require researching methodologies compatible with the software development life-cycle similar to the software transplantaton \cite{10.1145/3695987}.
	\item $RQ_2$ \textit{To what extent is software maintenance automated under the autopoietic \approach scheme powered by foundation models?} We can measure maintenance levels in a software system by considering corrective, adaptive, perfective, and preventive tasks executed by FMs agents.  
	\item $RQ_3$ \textit{Should software designs have a dedicated component to self-construction?} This question explores if and when incorporating \alife elements in software design. 
	\item $RQ_4$ \textit{How can we migrate from canonical maintenance to self-construction maintenance paradigm?} We can leverage software engineering research to migrate any potential system to an autopoietic architecture \approach.
\end{itemize}

To address the RQs, developing and deploying \approach will require extensive expertise in training and programming \textit{AI-agents}. An increased interest in AI-agents has emerged in recent years; this technology has enhanced the performance of challenging tasks\cite{yang2024sweagentagentcomputerinterfacesenable,wang2024executablecodeactionselicit,jimenez2024swebenchlanguagemodelsresolve, qiu2025alitagselfevolvinggenerativeagent}. We believe AI-agents theory and practice can contribute to the implementation of the proposed self-replicating components. For example, the control mechanism could be an AI agent trained on architectural patterns data. This control agent can also formulate questions to or communicate with the reasoning unit, another independent agent.

While the exact nature of inputs and outputs within autopoietic architectures \approach remains uncertain, the overarching goal is clear: to enable self-construction and maintenance through replication and adaptability processes inspired by \alife. It may be observed that the proposed architecture does not account for possible security issues or any other type of non-functional constraints. 




\section{Conclusion}
\label{sec:conclusion}



Software maintenance and evolution are challenging activities in \sdlc that require the application of Artificial Life (\alife) techniques combined with FM capabilities to advance next-generation artificial software engineering. This positional paper highlights the benefits of closing the gap between \alife and Software Engineering research by formulating \textit{autopoietic architectures} to enable an artificial self-construction software life-cycle. The autopoietic architecture, namely \approach, aims to follow Lehman laws for software evolution on a \sdlc (\ie continuing change, self-regulation, stability). \approach elicits the next generation of artificial software engineers for new \sdlc activities (\ie prompt-engineering, agentAI monitoring, and LLMs for SE debugging). Furthermore, \approach requires architectural data and benchmarking to assess the extent of self-construction of the original software system. 



\bibliographystyle{ACM-Reference-Format-num}
\bibliography{utils/references}


\end{document}